\documentclass[3p,times]{elsarticle}
\usepackage{graphicx,color}
\begin{document}
\title{Valence Band Electronic Structure of Pd Based Ternary Chalcogenide Superconductors}


\begin{frontmatter}




\title{Valence Band Electronic Structure of Pd Based Ternary Chalcogenide Superconductors}


\author[one,two]{H. Lohani}
\author[one]{ P. Mishra}
\author[three]{R. Goyal}
\author[three]{V. P. S. Awana}
\author[one,two]{B. R. Sekhar\corref{correspondingauthor}}
\ead{sekhar@iopb.res.in}
\address[one]{Institute of Physics, Sachivalaya Marg, Bhubaneswar 751005,
India.}
\address[two]{Homi Bhabha National Institute, Training School Complex, Anushakti
Nagar, Mumbai 400085, India}
\address[three]{National Physical Laboratory(CSIR),
Dr. K. S. Krishnan Road, New Delhi 110012, India.}
\begin{abstract}
We present a comparative study of the valence band electronic structure of Pd 
based ternary chalcogenide superconductors Nb$_2$Pd$_{0.95}$S$_5$, 
Ta$_2$Pd$_{0.97}$S$_6$ and Ta$_2$Pd$_{0.97}$Te$_6$ using experimental 
photoemission spectroscopy and density functional based theoretical 
calculations. We observe a qualitatively similarity between valence band (VB) 
spectra of Nb$_2$Pd$_{0.95}$S$_5$ and Ta$_2$Pd$_{0.97}$S$_6$. Further, we 
find a pseudogap feature in Nb$_2$Pd$_{0.95}$S$_5$ at low temperature, unlike 
other two compounds. We have correlated the structural geometry with the 
differences in VB spectra of these compounds. The different atomic packing in 
these compounds could vary the strength of inter-orbital hybridization among 
various atoms which leads to difference in their electronic structure as 
clearly observed in our DOS calculations.
\end{abstract}


\end{frontmatter}


\section{Introduction}
\label{1}
Recently discovered superconductors (SC) like Fe-pnictides\cite{stew}, 
Fe-chalcogenides\cite{taka}, SrRuO$_4$\cite{maeno} and organic 
SC\cite{norman,olsen}, driven by unconventional pairing have given a new 
impetus to the research in the field of superconductivity. Discovery of Pd 
based ternary chalcogenides, like Nb$_2$Pd$_{0.95}$S$_5$\cite{Jha,Zhang}, 
Nb$_2$PdSe$_5$\cite{khim}, Ta$_2$PdS$_5$\cite{Lu}, 
Ta$_2$Pd$_{0.97}$S$_6$\cite{Tiwari}, Ta$_2$Pd$_{0.97}$Te$_6$\cite{goyal} and 
Ta$_4$Pd$_3$Te$_{16}$\cite{jiao} is another advancement in this direction. 
These layered compounds provide a fertile ground for the existence of 
unconventional SC state owing to their Quasi-2-dimensional (Q2D) 
character\cite{pan,jiao}. 

One of the most interesting compounds in this low dimensional family, is 
Nb$_2$Pd$_{0.95}$S$_5$. It exhibits a SC transition 
temperature (T$_c$) around 6 K and a high value of Sommerfeld constant 
($\gamma$ = 32 mJ/mol-K$^2$) which signifies its strongly coupled SC nature. 
Resitivity of this compound shows a Fermi-liquid type behavior at low 
temperatures. Further, the heat capacity data is best fitted by using a two 
band model which is a signature of multiband superconductivity\cite{Jha}. 
Many unusual characters of these SCs have been analyzed from the point of 
view of spin-orbit scattering\cite{Lu,khim,Singh}, multiband 
effects\cite{miz,wolf} and unconventional pairing due to the presence of non 
centrosymmetric (NCS) structure\cite{Lu,youn,sigrist} etc. On the other hand, 
ternary compounds such as Ta$_2$Pd$_{0.97}$S$_6$ and 
Ta$_2$Pd$_{0.97}$Te$_6$ belonging to the same monoclinic structure (C$_{2/m}$) 
like Nb$_2$Pd$_{0.95}$S$_5$, show different behavior. The telluride compound 
Ta$_2$Pd$_{0.97}$Te$_6$, which crystallizes in Ta$_4$Pd$_3$Te$_{16}$ phase, 
has a much lower residual resistivity than normal state resistivity in 
comparison to the Nb$_2$Pd$_{0.95}$S$_5$. These compounds share a common 
structure composed of chains of Pd and Nb/Ta centred polyhedra with S/Te 
atoms. Changes in the structural geometry is a key factor determining the 
different physical behavior of these ternary SC.

Theoretical studies of the electronic structure of some of these materials, 
like Nb$_2$PdSe$_5$\cite{khim}, Nb$_2$PdS$_5$\cite{Zhang}, 
Ta$_2$PdS$_5$\cite{Singh}, Ta$_4$Pd$_3$Te$_{16}$\cite{David} have been 
reported recently and established the multiband nature of these compounds.
The Fermi surface (FS) of Nb$_2$PdS/Se$_5$, Ta$_2$PdS$_5$  compounds are composed 
of sheets of electron-hole character of different dimensions\cite{Singh,Zhang,David}, 
which could favor the  existence of various density wave instabilities, like charge density wave 
(CDW) and spin density wave (SDW)  due to the nesting between the 1-D like sheets of FS
in these systems\cite{David}. So far, there have been no reports on the experimental studies of the electronic 
structure of these compounds. In this paper, we discuss our photoemission 
results on Nb$_2$Pd$_{0.95}$S$_5$, Ta$_2$Pd$_{0.97}$S$_6$ and 
Ta$_2$Pd$_{0.97}$Te$_6$. We find that valence band(VB) spectra of 
Nb$_2$Pd$_{0.95}$S$_5$ and Ta$_2$Pd$_{0.97}$S$_6$ are qualitatively similar 
whereas it is quite different in case of Ta$_2$Pd$_{0.97}$Te$_6$. Further, we 
observe a pseudogap feature in Nb$_2$Pd$_{0.95}$S$_5$ at low temperature, 
unlike other two compounds. In these compounds the different atomic packing 
could change the strength of inter-orbital hybridization among various atoms
which leads to difference in their electronic structure as clearly seen in 
our DOS calculations. Our study highlights the role of structural geometry to
the different VB spectra of these compounds.

\section{Details of calculation}
\label{2}
Polycrystalline samples of Nb$_2$Pd$_{0.95}$S$_5$, Ta$_2$Pd$_{0.97}$S$_6$ and 
Ta$_2$Pd$_{0.97}$Te$_6$ were synthesized via solid state reaction route. 
Stoichiometry was confirmed by XRD measurements. Structural and other physical 
properties were studied and reported earlier\cite{Jha,Tiwari,goyal}. 
Photoemission spectra were collected in angle integrated mode using a 
hemispherical SCENTA-R3000 analyzer and a monochromatized He source 
(SCENTA-VUV5040). The photon flux was of the order of 10$^{16}$ 
photons/s/steradian with a beam spot of 2 mm in diameter. Fermi energy for 
all measurements were calibrated by using a freshly evaporated Ag film on to 
the sample holder. The total energy resolution, estimated from the width of 
the Fermi edge, was about 27 meV for the He I excitation. The  measurements 
were performed at a base pressure better than $\sim$ 3.0 $\times$ 10$^{-10}$ mbar.
 Samples were scrapped in a preparation chamber with a diamond file at base 
pressure 5.0 $\times$ 10$^{-10}$ mbar and the spectra were taken within an hour, 
so as to avoid any surface degradation. All the measurements were repeated 4-5 
times and observed almost identical spectra at each measurement.
For the temperature dependent measurements, the samples were cooled 
by pumping liquid nitrogen through the sample manipulator fitted with a 
cryostat. Sample temperatures were measured using a silicon diode sensor 
touching the bottom of the stainless steel sample plate. The low temperature 
photoemission measurements at 77 K were performed immediately after cleaning 
the sample surfaces followed by the room temperature measurements.

First-principles calculation were performed by using plain wave basis set 
inherent in Quantum Espresso (QE)\cite{qe}. Many electron exchange-correlation 
energy was approximated by Perdew-Burke-Ernzerhof (PBE) 
functional\cite{Ernzerhof,Wang,Chevary}. It was implemented in a scalar 
relativistic, ultrasoft pseudopotential\cite{Vanderbilt}. Fine mesh of 
k-points with Gaussian smearing of the order 0.0001 Ry was used for sampling 
the Brillouin zone integration and kinetic energy and charge density cut-off 
were set to 180 Ry and 900 Ry respectively. Experimental lattice parameters 
and atomic coordinates of Nb$_2$Pd$_{0.95}$S$_5$\cite{Jha}, 
Ta$_2$PdS$_6$\cite{philip} and Ta$_4$Pd$_3$S$_{16}$\cite{arthur}, after 
relaxed under damped (Beeman) dynamics with respect to both ionic coordinates 
and the lattice vector, were employed in the calculations. All parameters were 
optimized under several convergence tests.

\section{Results and discussion}
\label{3}
\begin{figure}
\includegraphics[width=7.5cm,keepaspectratio]{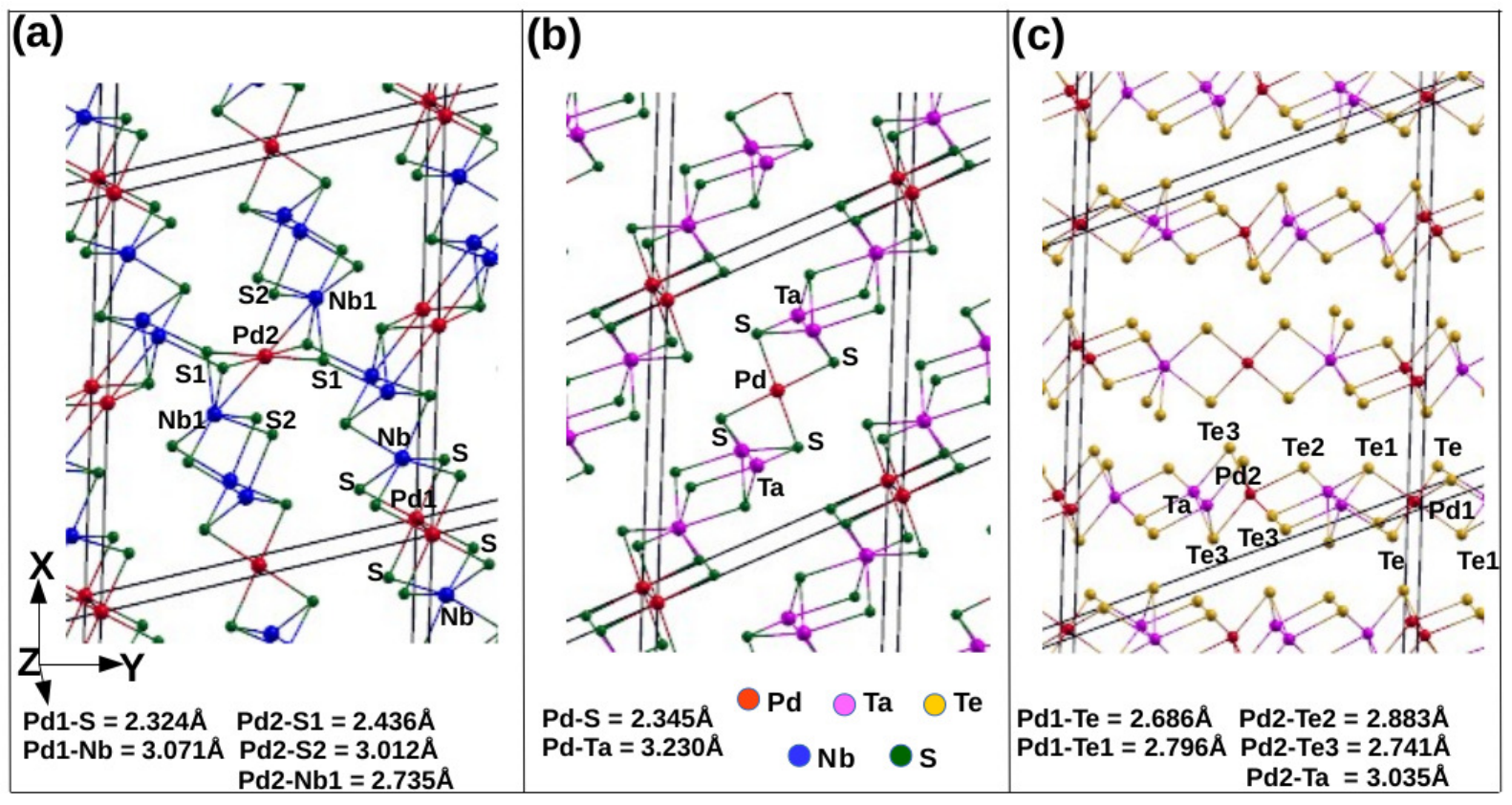}
\caption{\label{crystal}'Color online' (a), (b) and (c) are crystal structure 
of Nb$_2$PdS$_5$, Ta$_2$PdS$_6$ and Ta$_4$Pd$_3$Te$_{16}$ respectively. 
Different nearest neighbours of Pd atom at inequivalent sites and bond 
lengths are marked in each structure.}
\end{figure}

Fig.\ref{crystal}(a), (b) and (c) display the crystal structures of 
Nb$_2$PdS$_5$, Ta$_2$PdS$_6$ and Ta$_4$Pd$_3$Te$_{16}$ respectively. These 
layered compounds have a monoclinic structure, with space group symmetry 
C$_{2/m}$. The further details of crystal structures regarding lattice 
parameters, Wyckoff positions, site symmetry and fractional occupancies can 
be seen in Ref.[6], [27] and [28] for Nb$_2$PdS$_5$, Ta$_2$PdS$_6$ and 
Ta$_4$Pd$_3$S$_{16}$ respectively. These compounds exhibit complex structure 
consisting of chains of Nb/Ta and Pd centered polyhedra of S/Te atoms. In 
Ta$_2$PdS$_6$, Pd centered octahedra bridges two Ta centered trigonal 
prismatic polyhedra\cite{philip}. Similarly, Pd atoms have an edge sharing 
regular (Pd1) and edge sharing distorted (Pd2) octahedral coordination 
environment in Ta$_4$Pd$_3$Te$_{16}$\cite{arthur}. On the other hand, in 
Nb$_2$PdS$_5$, Pd atom has an octahedral coordination at Pd1 site and a 
distorted rhombohedral prismatic coordination at Pd2 site, which lies between 
the two adjacent layers\cite{douglas}. The coordination environment of Pd 
(Nb/Ta) atoms and their bond lengths between the various near neighbor (nn) 
S/Te vary in the three crystal structures which leads to different strengths 
of inter-orbital hybridization.

\begin{figure}
\includegraphics[width=7.5cm,keepaspectratio]{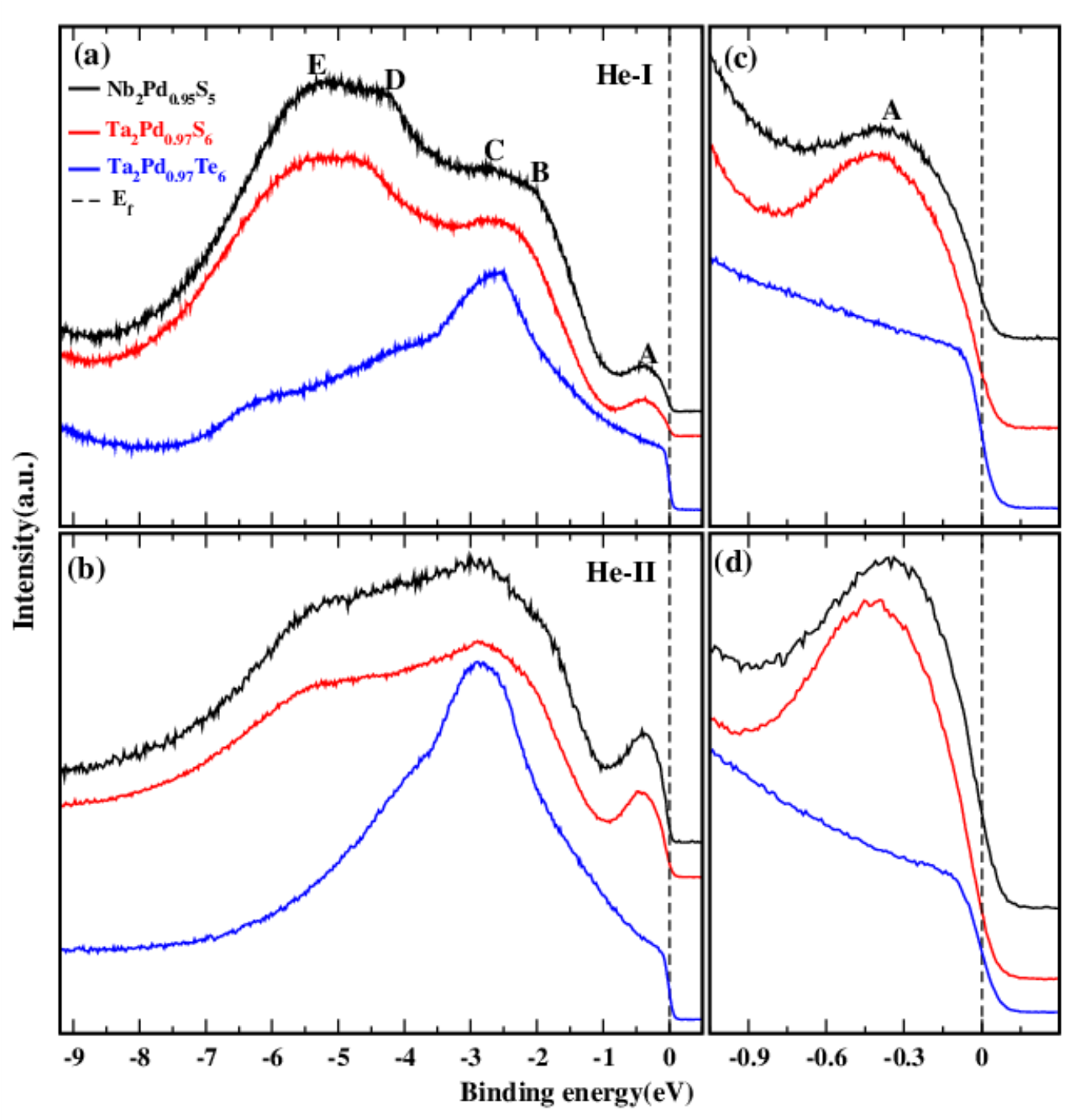}
\caption{\label{expval}'Color online' (a), (b) and (c) valence band spectra of
Nb$_2$Pd$_{0.95}$S$_5$(Black), Ta$_2$Pd$_{0.97}$S$_6$(Red) and 
Ta$_2$Pd$_{0.97}$Te$_6$(Blue)
obtained by using HeI and HeII excitation energy at 300 K respectively. 
(c) and (d) are an enlarge view of near E$_f$ region of the (a) and the (b) 
respectively.}
\end{figure}

Fig.\ref{expval}(a) and (b) show the valence band (VB) spectra of 
Nb$_2$Pd$_{0.95}$S$_5$, Ta$_2$Pd$_{0.97}$S$_6$ and Ta$_2$Pd$_{0.97}$Te$_6$ 
obtained by using HeI and HeII excitation energies at 300 K respectively. The 
spectra of Nb$_2$Pd$_{0.95}$S$_5$ show five features which include a hump at 
binding energy (BE) E$_b$ = -0.33 eV (A) and features at E$_b$ = -2.05 (B), 
-2.73 (C), -4.27 (D) and -5.29 (E) eV. In Ta$_2$Pd$_{0.97}$S$_6$, these 
features appear at higher BE. The hump structure A occurs at E$_b$ = -0.40 eV 
while feature B, C, D and E are observed at E$_b$ = -2.39, -2.84, -4.56 and 
-5.40 BE respectively. The VB spectra of Ta$_2$Pd$_{0.97}$Te$_6$ is markedly 
different from those of the other two compounds. The hump structure A and 
feature B are completely absent and a clear metallic edge is present at the 
Fermi Level (E$_f$). Furthermore, feature C, occurring at E$_b$ = -2.49 eV BE, 
appears as an intense peak structure while features D and E are faint. By 
tuning the photon energy from HeI (21.2 eV) to HeII (40.8 eV), a significant 
increase is observed in the spectral weight of the feature C in all the 
compounds as can be seen in Fig.\ref{expval} (b). While, the hump structure A 
becomes broader in Nb$_2$Pd$_{0.95}$S$_5$ and Ta$_2$Pd$_{0.97}$S$_6$, the 
metallic edge at E$_F$ in Ta$_2$Pd$_{0.97}$Te$_6$ appears smeared out slightly. 
These changes could be related to the different matrix elements involved in 
the photoemission process which depend on the ionization cross sections of 
atoms ($\sigma$). The $\sigma$\cite{lindau} multiplied by the number of 
valence electrons, for Pd$_{4d}$ to 
S$_{3p}$($\sigma_{Pd_{4d}}/\sigma_{S_{3p}}$) is 15 and 134 for HeI and HeII 
respectively. The $\sigma_{Nb_{4d}}/\sigma_{S_{3p}}$, 
$\sigma_{Ta_{5d}}/\sigma_{S_{3p}}$ and $\sigma_{Ta_{5d}}/\sigma_{Te_5p}$ 
increase by a factor of $\sim$ 2 for HeII with respect to HeI photon energy.
A comparison of the spectra taken with HeI and  HeII indicate that the 
enhanced feature C and hump A are mainly composed of Pd$_{4d}$ states. The 
relatively large intensification of feature C in Ta$_2$Pd$_{0.97}$Te$_6$ 
compared to the other two compounds could be due to the additional increment in 
$\sigma_{Pd_{4d}}/\sigma_{Te_{5p}}$ value by 20\% compared to 
$\sigma_{Pd_{4d}}/\sigma_{S_{3p}}$ for HeII. Similarly, composition of the 
metallic edge in Ta$_2$Pd$_{0.97}$Te$_6$ can be identified as mostly Te$_{5p}$ 
states. These changes are clearer in Fig.\ref{expval}(c) and (d) which 
displays an enlarged view of the near E$_f$ region for HeI and HeII 
respectively. The DOS at E$_f$ decreases in the order Ta$_2$Pd$_{0.97}$Te$_6$, 
Nb$_2$Pd$_{0.95}$S$_5$ and Ta$_2$Pd$_{0.97}$S$_6$ which agrees well with the 
trend in electrical resistivity shown by these 
compounds\cite{Jha,Tiwari,goyal}. Interestingly, observation of the sharp 
metallic edge in Ta$_2$Pd$_{0.97}$Te$_6$, unlike Nb$_2$Pd$_{0.95}$S$_5$, 
could also be the reason for the drastic reduction (65\%) in the ratio of 
normal state resistivity to the residual 
resistivity($\rho_{300 K}/\rho_{(0)}$) of Ta$_2$Pd$_{0.97}$Te$_6$ in 
comparison to Nb$_2$Pd$_{0.95}$S$_5$. 

\begin{figure}
\includegraphics[width=7.5cm,keepaspectratio]{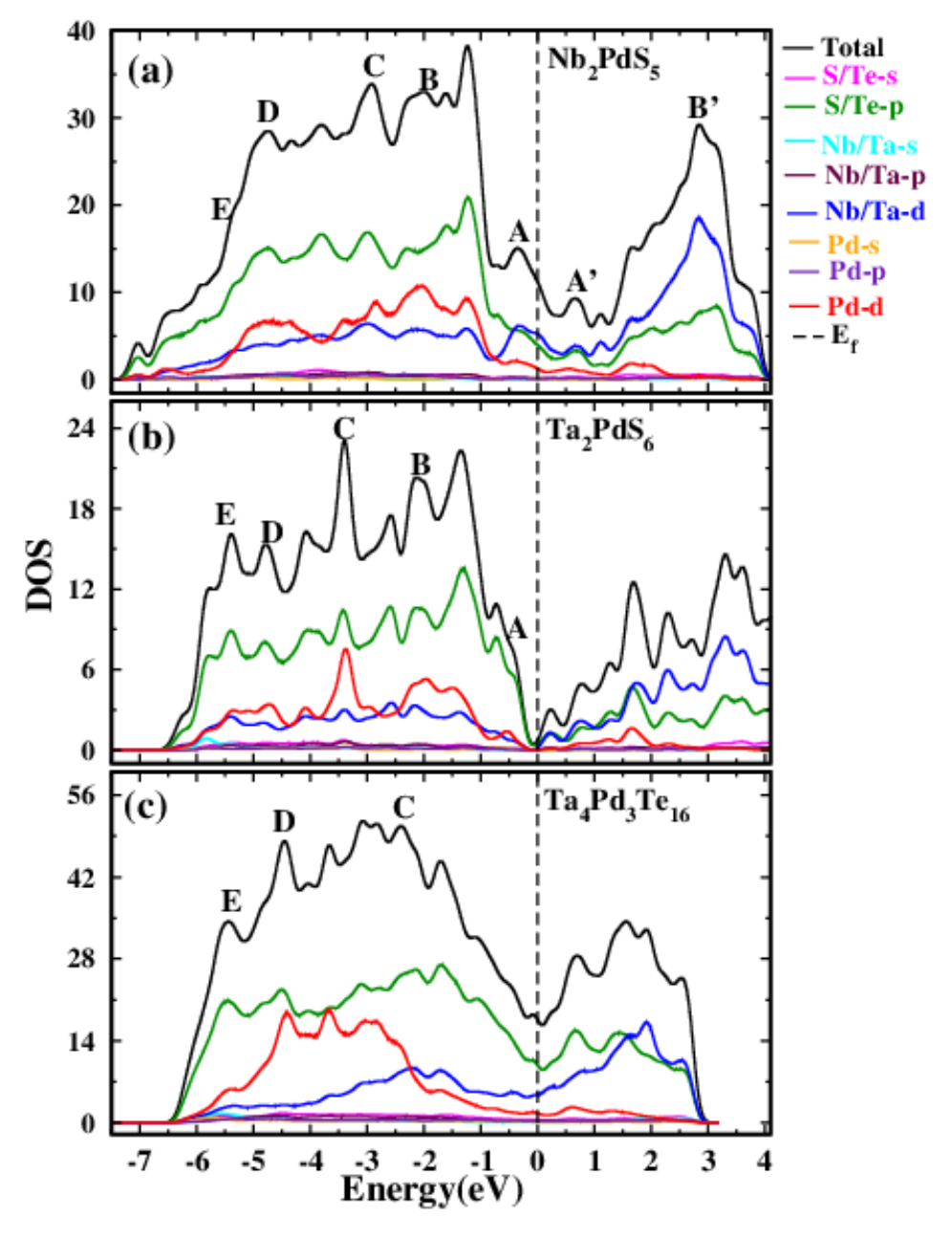}
\caption{\label{dos} 'Color online' (a), (b) and (c) show total calculated 
DOS with different atomic contributions of Nb$_2$PdS$_5$, Ta$_2$PdS$_6$ and 
Ta$_4$Pd$_3$Te$_{16}$ respectively.}
\end{figure}

In order to understand the origin of the VB features, we have performed 
electronic band structure calculations for similar but stoichiometric 
compositions; Nb$_2$PdS$_5$ and Ta$_2$PdS$_6$. We have used 
Ta$_4$Pd$_3$Te$_{16}$ composition to identify the VB features of 
Ta$_2$Pd$_{0.97}$Te$_6$ because it crystallizes in Ta$_4$Pd$_3$Te$_{16}$ phase 
as observed in the XRD measurements\cite{goyal}. In Fig.\ref{dos}(a), (b) and 
(c) calculated total DOS with various atomic contributions in cases of 
Nb$_2$PdS$_5$, Ta$_2$PdS$_6$ and Ta$_4$Pd$_3$Te$_{16}$ are shown 
respectively. In Nb$_2$PdS$_5$(Ta$_2$PdS$_6$) the conduction band (CB) states 
are dominated by S$_{3p}$-Nb$_{4d}$(Ta$_{5d}$) hybridized states while the VB 
(E$_b$ = -7.0 to -1.0 eV) comprises mainly of S$_{3p}$-Pd$_{4d}$ hybridized 
states in both the Nb$_2$PdS$_5$ and Ta$_2$PdS$_6$ compounds. In the 
near E$_f$ region sufficient states of S$_{3p}$-Nb$_{4d}$ hybridization are 
observed in Nb$_2$PdS$_5$ contrary to the situation in Ta$_2$PdS$_6$ where a 
dip can be seen in the DOS at the E$_f$. The gap like feature in 
Ta$_2$PdS$_6$ suggests that hybridization of Pd$_{4d}$ and Ta$_{5d}$ orbitals 
with their near neighbor (nn) S$_{3p}$ orbitals is stronger resulting in a 
clear separation between the bonding and antibonding states. Possibly this 
could be the reason for the occurrence of the features at higher BE in the 
experimental data of Ta$_2$Pd$_{0.97}$S$_6$ in comparison to the 
Nb$_2$Pd$_{0.95}$S$_5$ (Fig.\ref{expval}). On the other hand, in 
Ta$_4$Pd$_3$Te$_{16}$ the presence of Te atoms modifies the electronic 
structure to a great extent. These atoms are more distant from the Pd and Ta 
atoms which eventually reduces the strength of the inter-orbital hybridization 
of Pd$_{4d}$ and Ta$_{5d}$ with their nn Te$_{5p}$ orbitals. Therefore, 
Ta$_{5d}$ and Pd$_{4d}$ states are more localized than the same states in 
Nb$_2$PdS$_5$ and Ta$_2$PdS$_6$. Moreover, the large size of Te atom distorts 
the chain of Pd and Ta centered polyhedra and results in several 
Te-Te bonds\cite{David}. The hybridized states of these different Te-Te-5p 
orbitals diffuse within the VB and CB and the states crossing the E$_f$ 
acquire mostly the Te$_{5p}$ character. However, some small contribution from 
the Ta$_{5d}$ states are also observed in the vicinity of E$_f$ due to the 
weak Ta-Pd bond similar to the Nb-Pd bond at Pd2 site in 
Nb$_2$PdS$_5$(Fig.\ref{crystal}). The calculated DOS shows less intensity in 
the region between E$_f$ and E$_b$ = -2.0 eV in Ta$_4$Pd$_3$Te$_{16}$, unlike 
the other two compounds where many states (peak A and B) are observed in this 
region. This result qualitatively agrees with the absence of the hump A and 
feature B in the VB spectra of Ta$_2$Pd$_{0.97}$Te$_6$(Fig.\ref{expval}). 
Furthermore, various peaks are found in the calculated DOS which are not very 
clearly resolved in the VB spectra of all the compounds. However, the 
calculated  peaks, which are closer to the experimentally observed VB features
are marked accordingly(Fig.\ref{expval}). These results indicate that the VB 
features (B, C, and D) are mainly composed of S$_{3p}$-Pd$_{4d}$ hybridized 
states in Nb$_2$Pd$_{0.95}$S$_5$, and Ta$_2$Pd$_{0.97}$S$_6$ while in the case 
of Ta$_2$Pd$_{0.97}$Te$_6$ Pd-$_{4d}$ states are dominant in the higher BE 
region (feature C, and D) and the near E$_f$ states are dominated by 
Te$_{5p}$ (Fig.\ref{expval}). The identification of Pd$_{4d}$ character
of the experimental feature C on the basis of the intensity enhancement of 
this feature in HeII spectra(Fig.\ref{expval}) is consistent with the 
calculated DOS. The variation in the calculated DOS of these compounds 
supported by our experimental findings highlights the role of the different 
structural geometry which leads to different strengths in the inter-orbital 
hybridization of Pd and Nb/Ta with their nn S/Te atoms and thereby the 
electronic structure.

\begin{figure}
\includegraphics[width=7.5cm,keepaspectratio]{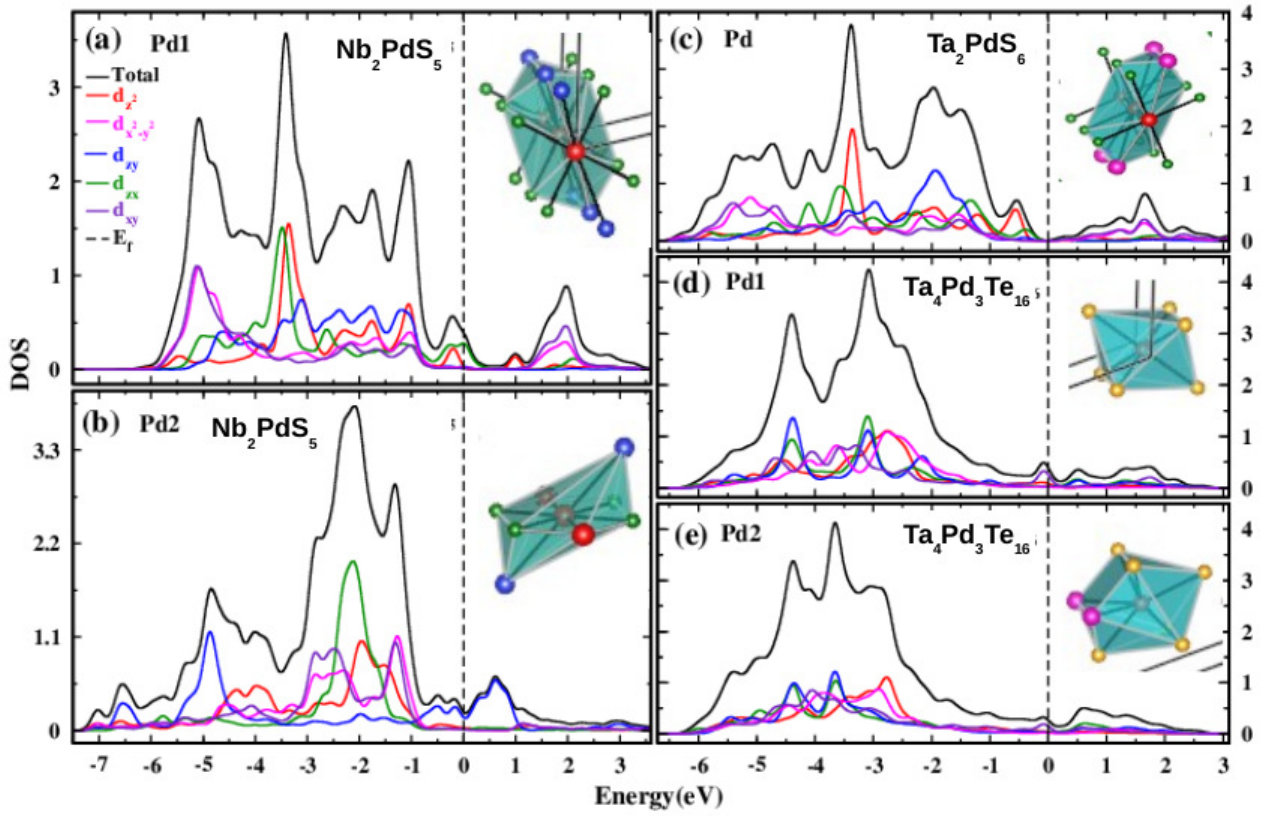}
\caption{\label{pddos}'Color online' DOS originated from different orbitals 
of Pd$_{4d}$ at site Pd1 and Pd2 are shown in (a) and (b) respectively in 
Nb$_2$PdS$_5$. (c) Pd$_{4d}$ DOS in Ta$_2$PdS$_6$. (d) and (e) show same DOS 
at Pd1 and Pd2 site respectively in Ta$_4$Pd$_3$Te$_{16}$. Inset of each plot 
shows coordination environment of each Pd atom in respective compound where
Cartesian axis are set in same orientation as shown in the crystal structure 
images(Fig.\ref{crystal}). Red, blue, green, magenta and yellow 
colour balls represent Pd, Nb, S, Ta and Te atoms respectively.}
\end{figure}

As discussed above, the VB of these ternary SCs are intimately linked to their 
complicated structural geometry. In order to investigate this issue in more 
detail, Pd$_{4d}$ m-projected DOS at differently coordinated sites are plotted 
in Fig.\ref{pddos}. Fig.\ref{pddos}(a) and (b) show the DOS of Pd$_{4d}$ at 
Pd1 and Pd2 site in Nb$_2$PdS$_5$ respectively. Fig.\ref{pddos}(c) shows DOS 
of Pd$_{4d}$ in Ta$_2$PdS$_6$. Fig.\ref{pddos}(d) and (e) show DOS of 
Pd$_{4d}$ at Pd1 and Pd2 site in Ta$_4$Pd$_3$Te$_{16}$ respectively. Effects 
of the strong inter-orbital hybridization with nn S$_{3p}$ orbitals are 
reflected as a large bandwidth of the Pd$_{4d}$ states. In Nb$_2$PdS$_5$, at 
the Pd1 site, 4d$_{x^2-y^2}$ and 4d$_{xy}$ orbitals lie in the plane of 
square planarly coordinated S atoms. Thus, these orbitals have the largest 
inter-orbital hybridization which leads to a clear separation between the 
bonding (-5.12 eV) and antibonding (2.00 eV) states originating from these 
orbitals. Similarly, the states derived from 4d$_{3z^2-r^2}$ and 4d$_{zx}$ 
orbitals which are directed towards nn Pd and Nb atoms, split into the bonding 
(-3.42 eV) and antibonding (-0.54 eV) states. However, in this case separation 
between the bonding and antibonding states is lesser due to the larger Pd-Pd 
(3.277\AA{}) and Pd-Nb(3.07\AA{}) inter-atomic distances in comparison to the 
smaller Pd-S(2.324\AA{}). Besides this, 4d$_{zy}$ states show a weakly 
hybridized nature. In contrast, at the Pd2 site, square planar coordination 
of the nn S atoms aligns in the YZ plane. Therefore, distinct bonding 
(-4.86 eV) and antibonding (0.64 eV) peaks are observed in 4d$_{zy}$ states. 
Similarly, 4d$_{3z^2-r^2}$, 4d$_{xy}$ and 4d$_{x^2-y^2}$ orbitals are weakly 
hybridized while the states constituted by 4d$_{zx}$ orbital show an atomic 
like character, localized at E$_b$ = -2.09 eV. The antibonding states of Pd2 
(4d$_{zy}$) is closer to E$_f$ compared to those of the Pd1 (4d$_{x^2-y^2}$ 
and 4d$_{xy}$) states and contribute mainly to the DOS at E$_f$. These 
differences could be linked to the relatively larger distance of nn S atoms at 
the Pd2 site (2.436\AA{}) than the Pd1 site (2.324\AA{}) which reduces the 
strength of inter-orbital hybridization. In Ta$_2$PdS$_6$, square planar 
coordination of Pd aligns in a plane identical to the Pd1 in Nb$_2$PdS$_5$ 
with a different orientation. Hence, the nature of DOS originating from 
different orbitals of Pd$_{4d}$ have similar character, like Pd1-4d DOS in 
Nb$_2$PdS$_5$.  On the other hand, in Ta$_4$Pd$_3$Te$_{16}$, for both Pd 
atoms (Pd1 and Pd2), the 4d$_{zy}$ and 4d$_{zx}$ states are almost degenerate 
and the energy difference among different 4d states is negligible, unlike the 
Nb$_2$PdS$_5$. These calculations show that the antibonding states originated 
from Pd2 (4d$_{zy}$) and Pd1 (4d$_{x^2-y^2}$ and 4d$_{xy}$) orbitals 
significantly contribute to the DOS at E$_f$ in Nb$_2$PdS$_5$, unlike the 
scenario in Ta$_2$PdS$_6$ and Ta$_4$Pd$_3$Te$_{16}$ where the Pd$_{4d}$ states 
are negligible at E$_f$. However, large number of Te-Te-5p hybridized states 
cross the E$_f$ in Ta$_4$Pd$_3$Te$_{16}$(Fig.\ref{dos}). Hence, the 
considerable states derived mainly from different Pd$_{4d}$/Nb$_{3d}$ and 
Te$_{5p}$ orbitals which are involved in crossing the E$_f$, is a signature 
for multiband effects in Nb$_2$PdS$_5$ and Ta$_4$Pd$_3$Te$_{16}$ respectively. 
This nature of multiband effects of different orbital character is consistent 
with previous reports\cite{Zhang,Singh,David}. Experimental realization of the 
multiband effects have also been claimed from the heat capacity behavior 
around T$_c$ in Nb$_2$Pd$_{0.95}$S$_5$\cite{Jha,Zhang}. This behaviour of 
orbital specific changes in the DOS of the Pd$_{4d}$ with different 
coordination environment, is reminiscent of the electronic behavior of 
Fe-based superconductors like FeSe and FeTe\cite{lohani}. In addition, the 
high contribution of Pd1 (4d$_{x^2-y^2}$ and 4d$_{xy}$) and Pd2 (4d$_{zy}$) 
DOS at E$_b$ = -5.12 eV and -2.09 eV in Nb$_2$PdS$_5$ are nearly close to the 
energy position of experimental VB features E and C in 
Nb$_2$Pd$_{0.95}$S$_5$(Fig.\ref{expval}) respectively. This could be a 
signature for the presence of two differently coordinated Pd atoms in 
Nb$_2$Pd$_{0.95}$S$_5$. 

\begin{figure}
\includegraphics[width=7.5cm,keepaspectratio]{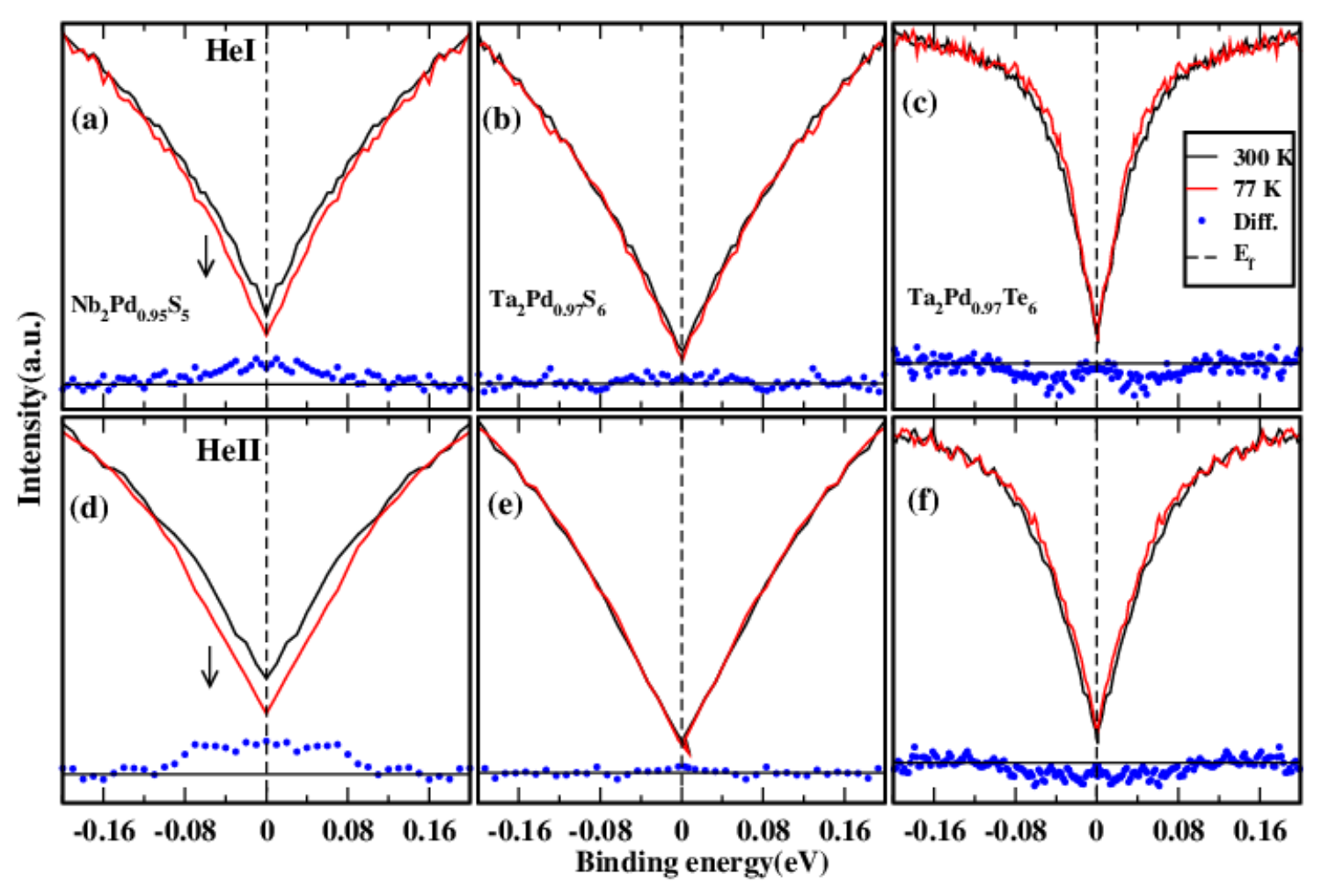}
\caption{\label{expfermi} 'Color online' Near E$_f$ VB spectra, at 300 K(Black) and 77 K(Red),
of Nb$_2$Pd$_{0.95}$S$_5$, Ta$_2$Pd$_{0.97}$S$_6$ and Ta$_2$Pd$_{0.97}$Te$_6$ collected with HeI energy
are shown in (a), (b) and (c) respectively. Same spectra collected at HeII energy
are shown in (d), (e) and (f) respectively. Blue dots represent difference between 
 the normalized  spectra  collected at 300 K and 77 K. Here, we have adopted the scheme of symmetrization 
to the spectra with respect to  E$_F$ in order to remove contribution arising
from Fermi-Dirac distribution function.}
\end{figure}

Fig.\ref{expfermi}(a), (b) and (c) represent the near E$_f$ VB spectra at 
300 K (Black) and 77 K (Red) of Nb$_2$Pd$_{0.95}$S$_5$, 
Ta$_2$Pd$_{0.97}$S$_6$ and Ta$_2$Pd$_{0.97}$Te$_6$ respectively taken using 
the HeI source. Fig\ref{expfermi}(d), (e) and (f) depict the same spectra at 
300 K and 77 K for the HeII excitation energy. Here, we have adopted the 
scheme of symmetrization\cite{ber} to the spectra with respect to the E$_f$ 
in order to remove contribution arising from the Fermi-Dirac distribution 
function. In order to highlight the temperature dependent changes, differences 
between the normalized spectra collected at 300 K and 77 K are also plotted 
in each graph. In case of Nb$_2$Pd$_{0.95}$S$_5$, states in the vicinity of 
E$_f$ are suppressed at 77 K in comparison to those at 300 K. This indicates 
the opening of a small($\Delta$ =  95 meV ) pseudogap which is more clear in 
the spectra taken with HeII. In this sample a crossover has been observed 
from electron to hole dominated transport carriers below 100 K in the Hall 
measurements\cite{Jha,yang}. This observation along with the presence of a 
pseudogap is again consistent with the pseudogap driven sign reversal in Hall 
coefficient predicted by Evtushinsky{\it et al.}\cite{foll}. The coupling 
strength(2$\Delta$/k$_B$T) estimated at this crossover temperature (100 K) 
from the gap value ($\Delta$) is anomalously large ($\sim$ 20) which 
indicates little possibility for any density wave instabilities, like charge 
density wave (CDW) and spin density wave (SDW) in the system to form the 
pseudogap. In addition, importance of electronic correlations have also not 
been found in earlier studies on this compound\cite{Jha,khim,Zhang}. On the 
other hand, the near E$_f$ states remain unchanged with the lowering of 
temperature to 77 K in Ta$_2$Pd$_{0.97}$S$_6$ and Ta$_2$Pd$_{0.97}$Te$_6$. 
Therefore, the origin of the observed pseudogap in Nb$_2$Pd$_{0.95}$S$_5$ 
could be associated to its complex structural geometry.

\section{Conclusion}
We have investigated the valence band electronic structure of Pd based ternary 
chalcogenide superconductors Nb$_2$Pd$_{0.95}$S$_5$, Ta$_2$Pd$_{0.97}$S$_6$ 
and Ta$_2$Pd$_{0.97}$Te$_6$ using experimental valence band photoemission and 
theoretical band structure calculations. We observe that the VB spectra of 
Ta$_2$Pd$_{0.97}$S$_6$ is qualitatively similar to Nb$_2$Pd$_{0.95}$S$_5$
except slight differences in the energy position of various features. On the 
other hand, the VB spectra of Ta$_2$Pd$_{0.97}$Te$_6$ differs remarkably from 
the other two compounds, particularly in the near E$_f$ region where a clear 
metallic edge is observed, unlike the other compounds. Our study also shows 
the existence of a temperature dependent pseudogap in Nb$_2$Pd$_{0.95}$S$_5$.
whereas, the near E$_f$ states remain unchanged with the lowering of 
temperature to 77 K in Ta$_2$Pd$_{0.97}$S$_6$ and Ta$_2$Pd$_{0.97}$Te$_6$. 
These changes seen the VB spectra could be due to their different structural 
geometry which provides different strengths to the inter-orbital hybridization 
of Pd and Nb/Ta with their nn S/Te atoms. This modifies the electronic 
structure significantly as clear from the plots of the calculated DOS of these 
compounds. The calculated DOS shows that the VB features of Nb$_2$PdS$_5$ and 
Ta$_2$PdS$_6$ are mainly comprised of the S$_{3p}$-Pd$_{4d}$ hybridized 
states while the metallic edge in Ta$_2$Pd$_{0.97}$Te$_6$ is predominant with 
Te-$_{5p}$ character. Our comprehensive study provides a deeper insight into 
the VB states of these Pd based ternary compounds in correlation with their 
different structural geometry.


\begin{thebibliography}{200}
\bibitem{stew} G. R. Stewart {\it Rev. Mod. Phys.} {\bf 83} (2011) 1589.
\bibitem{taka} Y. Mizuguchi and Y. Takano {\it J. Phys. Soc. Jpn.} {\bf 79} (2010) 102001.
\bibitem{maeno} A. P. Mackenzie, Y. Maeno {\it Rev. Mod. Phys.} {\bf 75} (2000) 657.
\bibitem{norman} M. R. Norman {\it Science} {\bf 332}  (2011) 196.
\bibitem{olsen} K. Bechgaard, K. Carneiro, M. Olsen, F. B. Rasmussen, C. S. Jacobsen {\it Phys. Rev.
Lett.} {\bf 46} (1981) 852.
\bibitem{Jha} R. Jha, B. Tiwari, P. Rani, H. Kishan and V. P. S. Awana { \it J. Appl. Phys.} {\bf115} (2014) 213903.  
\bibitem{Zhang} Q. Zhang, G. Li, D. Rhodes, A. Kiswandhi, T. Besara, B. Zeng, J. Sun, Siegrist, M. D. Johannes and L. Balicas 
{\it Scientific Reports} {\bf 3} (2013) 1446.
\bibitem{khim} Seunghyun Khim, Bumsung Lee, Ki-Young Choi1, Byung-Gu Jeon, Dong Hyun Jang, Deepak Patil, Seema Patil,
Rokyeon Kim, Eun Sang Choi, Seongsu Lee, Jaejun Yu and Kee Hoon Kim { \it New J. Phys.} {\bf 15} (2013) 123031.
\bibitem{Lu} Y. F. Lu, T. Takayama, A. F. Bangura, Y. katsura, D.     Hashizume and  H. Takag {\it J. Phys. Soc. Japan}
{\bf 83} (2014) 023702.
\bibitem{Tiwari} B. Tiwari, B. B. Prasad, R. Jha, D. K. Singh and V. P.S. Awana  { \it J. of Super.  Nov. Magn.} {\bf 27} (2014) 2181.
\bibitem{goyal} R. Goyal, B. Tiwari,  R. Jha,  and V. P.S. Awana  { \it J. of Super.  Novel. Magn.} {\bf 28} (2015) 1195.
\bibitem{jiao} Wen-Hu Jiao, Zhang-T Tang, Yun-Lei Sun, Y. Liu, Q. Tao, Chun-Mu Feng, Yue-Wu Zeng, Zhu-An Xu
and Guang-Han Cao {\it J. of Am. Chem. Soc.} {\bf 136} (2014) 1284.
\bibitem{pan} J. Pan, W. H. Jiao, X. C. Hong, Z. Zhnag, L. P. He, P. L. Cai, G. H. Cao and S. Y. Li
{\it Phys. Rev. B} {\bf 13} (1976) 3284.
\bibitem{chand}  B. S. Chandrasekhar {\it Appl. Phys. Lett.} {\bf 1} (1962) 7.
\bibitem{clog}  A.M. Clogston {\it Phys. Rev. Lett.} {\bf 9} (1962) 266.
\bibitem{Singh} David J. Singh { \it Phys. Rev. B} {\bf 88} (2013) 174508.
\bibitem{miz}Masahiro Takahashi, Takeshi Mizushima and Kazushige Machida { \it Phys. Rev. B} {\bf 89} (2014) 064505.
\bibitem{wolf}D.A. Zocco, K. Grube, F. Eilers, T. Wolf and H.v. Lohneysen {\it Phys. Rev. Lett.} {\bf 111} (2013) 057007.
\bibitem{youn} Suk Joo Youn, Mark H. Fischer, S. H. Rhim, Manfred Sigrist, and Daniel F. Agterberg
{ \it Phys. Rev. B} {\bf 85} (2012) 220505(R)
\bibitem{sigrist} Jun Goryo, Mark H. Fischer and Manfred Sigrist { \it Phys. Rev. B} {\bf 86} (2012) 100507(R)
\bibitem{David} David J. Singh { \it Phys. Rev. B} {\bf 90} (2014) 144501.
\bibitem{qe} Giannozzi P. et al.  {\it http:/www.quantum-espresso.org}. 
\bibitem{Ernzerhof}
Perdew J P,Burke K and Ernzerhof M  {\it Phys. Rev. Lett.} { \bf 77} (1996) 3865.
\bibitem{Wang} Perdew J P and Wang y  { \it Phys.Rev.B} { \bf 45} (1992) 13244.
\bibitem{Chevary} Perdew J P, Chevary J A, Vosko S H, Jackson K A, Pederson M R, Singh D J and Fiolhais C  {\it Phys.Rev. B} {\bf 46} (1992) 6671.
\bibitem{Vanderbilt}
Vanderbilt D  Soft self-consistent pseudopotential in a generalized eigenvalue formalism {\it Phys.Rev.B} {\bf 41} (1990) 7892.
\bibitem{philip} D. A. Keszler, P. J. Squattrito, N. E. Brese, J. A. Ibers, S. Maoyu and L. Jiaxi {\it Inorg. Chem.} {\bf 24} (1985) 3063. 
\bibitem{arthur}  A. Mar and J. A. Ibers {\it J. Chem. Soc. Dalton Trans.} {\bf 1} (1991) 639. 
\bibitem{douglas} D. A. Keszler, J. A. Ibers, S. Maoyu and L. Jiaxi {\it J. of Solid State Chem.} {\bf 57} (1985) 68.
\bibitem{lindau} J.J. Yeh and I.Lindau {\it Atomic Data and Nuclear Data Tables} {\bf 32}  (1985) 1-155.
\bibitem{lar} P. Larson, V. A. Greanya, W. C. Tonjes, R. Liu, S. D. Mahanti and C. G. Olson { \it Phys. Rev. B} {\bf 65} (2002) 085108.
\bibitem{lohani} H. Lohani, P. Mishra and B. R. Sekhar {\it Physica C} {\bf 512} (2015) 54.
\bibitem{ber}S. V. Borisenko, A. A. Kordyuk, A. N. Yaresko, V. B. Zabolotnyy, D. S. Inososv, R. Schuster,
B. Buchner, R. Weber, R. Follath, L. Patthey and H. Berger {\it Phys. Rev. Lett.} {\bf 100} (2008) 196402.
\bibitem{yang} X. Ding, Y. Pan, H. Yang and Hai-Hu Wen { \it Phys. Rev. B} {\bf 89} (2014) 224515.
\bibitem{foll}D. V. Evtushinsky, A. A. Kordyuk, V. B. Zabolotnyy, D. S. Inosov, B. Buchner, H. Berger, L. Patthey,
R. Follath and  V. Borisenko  {\it Phys. Rev. Lett.} {\bf 100} (2008) 236402.
\bibitem{gabo}Alexander Gabovich, A I Voitenko and Toshikazu Ekino {\it J. Phys. Condens Matter} {\bf 16} (2004) 3681.
\end{thebibliography}
\end{document}